\documentclass[global,twocolumn]{svjour}
\journalname{Applied physics B}
\usepackage{graphics}
\usepackage{hyperref}
\usepackage{amsmath}
\usepackage[T1]{fontenc}
\DeclareTextSymbol{\degre}{T1}{6}
\DeclareTextSymbol{\degre}{OT1}{23}

\begin{document}

\title{AOPDF-shaped optical parametric amplifier output in the visible}

\author{A. Monmayrant\inst{1}, A. Arbouet\inst{1}, B. Girard\inst{1}, B. Chatel\inst{1} \and A. Barman\inst{2}, B. J. Whitaker\inst{2}
and D. Kaplan\inst{3}}

\institute{Laboratoire Collisions, Agr\'egats R\'eactivit\'e (CNRS UMR 5589), IRSAMC Universit\'e Paul Sabatier, 31062
Toulouse, France \and School of Chemistry, University of Leeds,
Leeds, LS2 9JT, UK \and Fastlite, Batiment 403, Ecole Polytechnique, Palaiseau, France}

\maketitle

\begin{abstract}Time Shaping of ultrashort visible pulses has been performed using
a specially designed Acousto-Optic Programmable Dispersive Filter
of 50\% efficiency at the output of a two-stage noncollinear
optical parametric amplifier. The set-up is compact and reliable.
It provides a tunable shaped source in the visible with unique
features: 4 ps shaping window with preserved tunability over
500-650 nm, and pulses as short as 30 fs. Several $\mu$J output
energy is easily obtained.\end{abstract}



\noindent

The development of femtosecond laser technology has opened access to
unforeseen applications in molecular, chemical physics, as well as
biology \cite{Assion98,RabitzScience2000,Levis01,Motzkus02bio}. In
the past ten years, the active control of ultrafast physical or
chemical processes by means of well-defined shaped laser pulses has
become possible \cite{Weinacht99,Degert02CTshaped}. The high number
of potential applications of femtosecond pulse shaping turned it
very quickly into a very intense field of research.

Traditional methods for femtosecond pulse shaping are based on a
Liquid Crystal Device (LCD) or Acousto-Optic Modulator (AOM) placed
in the Fourier plane of a grating based zero dispersion 4f
configuration \cite{weiner00,stobrawashap01,pulseshaperRSI04}. The
different wavelengths are spatially separated and can then be
addressed individually. Spectacular results have been obtained with
such devices \cite{Goswami03Review,Dantus04review}. However, the
pixelation causes pre- and post-pulses in the time domain containing
sometimes a substantial fraction of the total pulse energy
\cite{WeferNelson95slm}. Changing the wavelength requires careful
realignment, thus precluding easy tunability. Finally, their large
size can be a severe limitation in some applications.

Visible shaped pulses have been obtained by inserting such pulse
shapers in a Noncollinear Optical Parametric Amplifier (NOPA)
\cite{ZeidlerAPB02,SchreiberOL01}. Either the white light seed was
shaped with a LCD \cite{ZeidlerAPB02}, or the output of the first
stage of a two-stage NOPA was shaped with an AOM
\cite{SchreiberOL01}. Despite significant results, the complexity of
the set-up was increased dramatically and little tunability was
available. Also the constraint of temporal overlap between shaped
and pump pulses reduced the shaping window to ca 200 fs
\cite{ZeidlerAPB02} or 1 ps \cite{SchreiberOL01}. In other studies,
the 4f shaper was placed directly at the output of a visible laser
\cite{Motzkus02bio,ReitzeAPL1992,XuIEEE2000,Motzkus00Adaptativecompression}.
A much simpler scheme, avoiding the complexity of the 4f line, and
with higher efficiency, can be achieved using instead an
Acousto-Optic Programmable Dispersive Filter (AOPDF)
\cite{verluise00} at the output of a NOPA.

AOPDF are based on the propagation of light in an acousto-optic
birefringent crystal. The interaction of an incident ordinary
optical wave with a collinear acoustic shear wave leads to
diffraction of an extraordinary wave. Spectral phase and amplitude
pulse shaping of a femtosecond optical pulse can be achieved by
controlling the amount of extraordinary versus ordinary propagation
in the optical path of each of its spectral components.
The collinear acousto-optic interaction and the reduced size result
in an easy-to-align device, appropriate for insertion in an
amplified laser chain or in a pump-probe setup.
AOPDF have proven to be very useful to correct the time aberrations
introduced in Chirped Pulse Amplifiers, for amplitude and phase
control of ultrashort pulses \cite{verluise00,Pittman02}, or even in
characterization set-ups \cite{Monmayrant_OL_2003_28}.

Most of AOPDF applications have been restricted so far to the near
infrared. Indeed, phase-matching conditions for shorter wavelengths
require higher acoustic frequencies for which absorption is
increased. Actually, preliminary experiments demonstrated that
acoustic absorption is predominant in the blue region of the
spectrum \cite{Kaplan_up_2004}. To overcome these limitations, we
have specially designed a new AOPDF accepting pulse energy densities
up to 300 $\mu$J/$\textrm{cm}^2$ with reduced absorption at
wavelengths as low as 500 nm.


Here, we report on pulse compression and shaping directly at the
output of a home made NOPA with this new AOPDF. This results in a
simple, compact, and reliable device providing sub-30fs pulses on a
4 ps shaping window, easily tunable
in the 500-650 nm range.

The design of the AOPDF must fulfill two requirements: the optical
yield has to be maximum because any loss will not be recovered in
further amplification stages and the shaping capabilities of the
device must be preserved. In particular, the device should
compensate for its own basic dispersion (due to the wavelength
dependent refractive index). Otherwise, obtaining the shortest
pulses will require an additional compressor device, undermining the
simplicity of the approach.

The acoustic beam orientation being set to align the group
velocity of both acoustic and optical beams, the main design
parameter is the angle of propagation of the latter, $\theta =
([110], k)$, in the birefringent crystal $\textrm{Te0}_2$.

The diffraction efficiency (i.e. the fraction of the energy at a
given optical wavelength recovered in the diffracted beam) has a
maximum at $\theta = 58.5$\degre and decreases at smaller angles
\cite{Kaplan02}.

The maximum programmable delay for an incoming pulse centered at
$\lambda_{opt}$ is:
\begin{equation}
 T_{max} =  \Delta n_g (\lambda_{opt}) cos^2
(\theta) L/c
\end{equation}
where
\begin{equation}
\Delta n_g (\lambda_{opt}) = n_{g,e} (\lambda_{opt}) - n_{g,o}
(\lambda_{opt})
\end{equation}
 is the group
birefringence, L the length of the crystal and c the speed of light. Part of this delay capacity, called
$T_{\textrm{comp}}$ , will be used to compensate the dispersion of the device and the remaining delay
$T_{\textrm{max}}-T_{\textrm{comp}}$ available for pulse shaping is a decreasing function of $\theta$.

The acoustic absorption $\beta$ in the crystal has a quadratic
dependence on the acoustic frequency $f$ and a non-trivial
dependence on $\theta$ \cite{Woodruff_Ehrenreich}:

\begin{equation}
\beta = \mathcal{C} \frac{\gamma^2 (\theta) f^2(\theta )}{V^4
(\theta)}
\end{equation}
$\mathcal{C}$ is a constant, $V$ is the acoustic phase velocity, and
$\gamma$ the Gruneisen constant. A $\beta$ value of 18
dB/$\mu$s.GHz$^2$ has been measured experimentally \cite{Kaplan02}
for $\theta = 0$\degre.
Although a precise determination of the $\gamma(\theta)$ law is
difficult, one can infer that $\beta$ decreases with $\theta$.

\begin{figure}[htb]
\resizebox{0.48\textwidth}{!}{\includegraphics{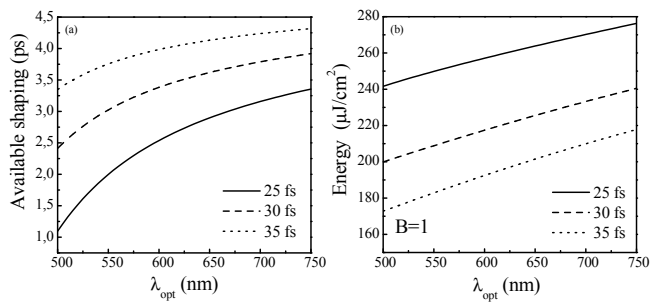}}
\caption{Pulse shaping capability (a) and intensity capability (b)
of the AOPDF for different FWHM Fourier limited pulse durations, as
a function of wavelength.}
\end{figure}

Taking into account these 
constraints, a 25 mm crystal at $\theta = 45$\degre has been
designed giving excellent results in terms of shaping capacities and
optical output power as described below. The computed temporal width
of the shaping window is depicted in Fig. 1a as a function of the
optical wavelength for several FWHM pulse durations. The AOPDF
allows shaping on up to 4 ps. The propagation of 30 fs pulses
through the whole AOPDF yields B integrals (accumulated self phase
modulation) of unity for intensities above 200
$\mu$J/$\textrm{cm}^2$ on the 500-750 nm range (Fig. 1b).

Experiments have been performed using this newly designed AOPDF at
the output of a two-stage NOPA
\cite{chatelNa03PRA,RiedleAPB00NOPAVisNIR}. The output beam of the
NOPA was split into two parts as shown in Fig. 2. One part was
compressed in a silica prism compressor, leading to durations around
25-30 fs on the 500-650 nm range. This value is close to 20 fs which
is the Fourier Transform limited pulse duration corresponding to the
FWHM of the intensity spectrum. The other part remained uncompressed
and was fed directly in the AOPDF. The beam profile is adjusted so
that its FWHM lies from 1.5 to 2.5 mm. Using a variable neutral
density filter, the energy in front of the AOPDF was varied from 1
to 6 $\mu$J, corresponding to energy densities up to 300
$\mu$J/$\textrm{cm}^2$ without any major self phase modulation
effect, a value somewhat higher than the theoretical one. Pulses
coming out of the AOPDF are characterized either by 2nd order
autocorrelation or by cross-correlation with the compressed
reference beam (see Fig. 2).

\begin{figure}[htb]
\resizebox{0.48\textwidth}{!}{\includegraphics{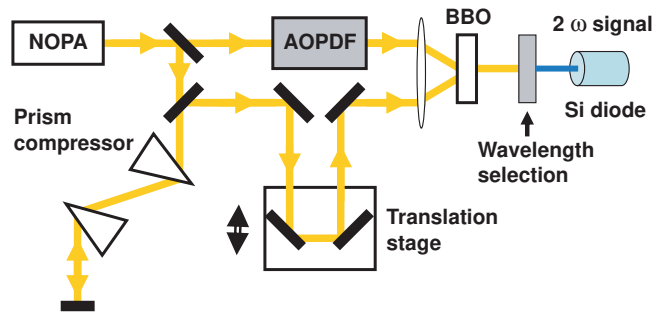}}
 \caption{Experimental setup. NOPA output 12 $\mu$J
(pumped with 800 nm, 120 fs, 250 $\mu$J pulses).  }
\end{figure}

In a first series of experiments, acoustic waves allowing
compensation of the quadratic phase term including the pulse initial
chirp ($\sim 1000 \textrm{fs}^2$) and $\textrm{Te0}_2$ induced chirp
($\sim 25000 \textrm{fs}^2$) together with higher order phase terms
have been programmed. This lead to compression of the output of the
NOPA down to sub-30 fs pulses on the 500-650 nm range.

Fig. 3 shows the spectrum after the AOPDF at 510, 550 and 640 nm
together with second harmonic intensity autocorrelation in a 100
$\mu$m BBO. The FWHM is sub-30 fs assuming a $\textrm{sech}^2$ pulse
profile. Additional experiments performed using a similar set-up
where the NOPA is a commercial Clark-MXR NOPA confirmed that the
small pedestal at 510 nm in Fig. 3 could be reduced. The optical
yield of the AOPDF is commonly 50 \% on the whole spectral range. 75
\% can be achieved still leading to pulses in the 30 fs range but
with less shaping capacity (all results below have been obtained
with 50 \% optical yield). However, at these high throughputs,
saturation of the acoustic wave should be carefully controlled.

\begin{figure}[htb]
\resizebox{0.48\textwidth}{!}{\includegraphics{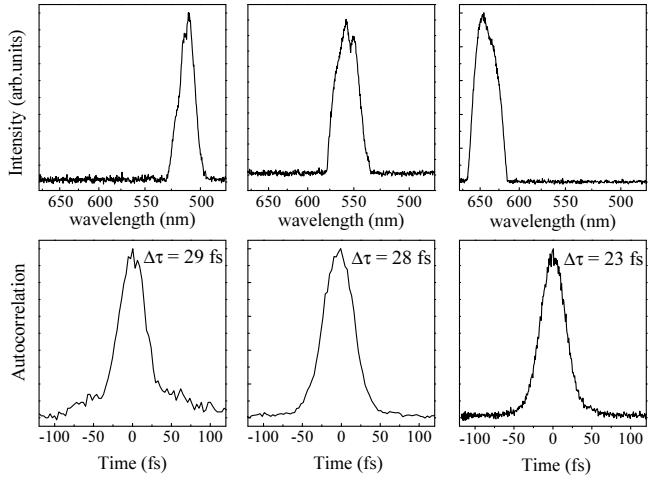}}
\caption{Spectrum and intensity autocorrelation for various
wavelengths, showing sub-30 fs pulses. Spectra have been displayed
in reciprocal scale to allow direct comparison of the FWHM between
the various center wavelengths.}
\end{figure}

The AOPDF is not only able to compress the pulse over the whole
spectral range but also to shape it in amplitude and phase. Fig. 4
shows several cross-correlations performed in a 20 $\mu$m thick
BBO crystal between the output of the NOPA compressed by prisms
and the output of the AOPDF. The cross-correlation signal as a
function of the delay in the AOPDF is shown in Fig. 4a. The zero
delay corresponds to a pulse diffracted in the middle of the
crystal and positive delays to pulses diffracted on the input side
of the AOPDF. From -0.5 ps to 2.5 ps the amplitude of the signal
is constant demonstrating the capacity of the AOPDF to generate
delays up to 3 ps without attenuation and even 4 ps with a
moderate attenuation in accordance with the computed values of
Fig. 1b. For delays between -0.5 and -1.5 ps, the
cross-correlation intensity decreases as a result of acoustic
absorption: the optical beam is diffracted by an acoustic pulse
that propagates along a longer path in the crystal. Multiple
output pulses have been generated, simply summing multiple
acoustic pulses, on the whole spectral range. An example of a
5-pulse sequence is shown Fig. 4b at 640 nm. Quadratic phase up to
$4.10^4$ $\textrm{fs}^2$ or cubic phase up to $3.10^5 $
$\textrm{fs}^3$ have been also programmed successfully.

\begin{figure}[h]
\resizebox{0.48\textwidth}{!}{\includegraphics{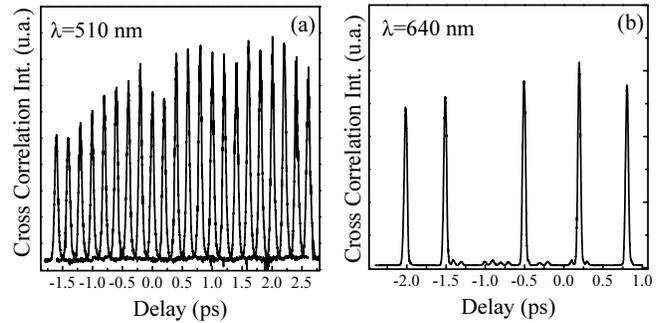}}
\caption{Cross-correlation signal between the NOPA output compressed
by prisms and the output of the AOPDF. (a) Several delays obtained
every 200 fs at $\lambda$ = 510 nm. (b) Multiple pulses at $\lambda$
= 640 nm.}
\end{figure}

 In this paper, the design of an AOPDF crystal
optimized for shaping in the visible and the results of pulse
compression and shaping experiments are presented.
Optical yield up to 50 \% have been obtained with input energies of
6 $\mu$J and an energy density of 300 $\mu$J/$\textrm{cm}^2$ without
any major self phase modulation phase effect, on a never reached
(for this spectral range) temporal window of 4 ps. Therefore, the
AOPDF appears to be perfectly appropriate for tailoring the output
of a NOPA with typical specifications of 15 $\mu$J energy and 2.5 mm
diameter FWHM. Shorter durations or higher energies can be obtained
by stretching the pulse before feeding the AOPDF and using an
external compressor \cite{KaplanKrausz03}. To remain below the
$\textrm{Te0}_2$ crystal damage threshold, an option would be to
magnify the beam diameter provided that the acoustic beam inside the
crystal has been scaled up. The performances in terms of maximum
energy, optimal compression and temporal shaping window make it an
ideal tool for tunable wavelength pulse shaping in the visible. Its
large tunability (500-650 nm) and broad temporal window, together
with its high update rate, makes this new device a unique tool for a
feedback loop in optimal control experiments
\cite{Assion98,RabitzScience2000,Levis01,Motzkus02bio,Goswami03Review,Dantus04review,Warren93}.
The authors thank Pierre Tournois for fruitful discussions.

\bibliographystyle{osajnl}
\bibliography{biblioAOPDF}

\end{document}